\documentclass[aps,prl,a4paper,showpacs,twocolumn,floatfix]{revtex4}
\usepackage[latin1]{inputenc}
\usepackage{amsfonts}
\usepackage{amssymb}
\usepackage{amsmath}
\usepackage{graphicx}
\usepackage{epsfig}
\usepackage{bm}
\usepackage{ulem}

\begin{document}

\title{Temporally multiplexed quantum repeaters with atomic gases}
\date{\today}
\author{Christoph Simon$^1$, Hugues de Riedmatten$^2$, and Mikael Afzelius$^2$}
\affiliation{$^1$ Institute for Quantum Information Science
and Department of Physics and Astronomy, University of
Calgary, Calgary T2N 1N4, Alberta, Canada\\$^2$ Group of
Applied Physics, University of Geneva, CH-1211 Geneva 4,
Switzerland}

\begin{abstract}
We propose a temporally multiplexed version of the Duan-Lukin-Cirac-Zoller (DLCZ) quantum repeater protocol using controlled inhomogeneous spin broadening in atomic gases. A first analysis suggests that the advantage of multiplexing is negated by noise due to spin wave excitations corresponding to unobserved directions of Stokes photon emission. However, this problem can be overcome with the help of a moderate-finesse cavity which is in resonance with Stokes photons, but invisible to the anti-Stokes photons. Our proposal promises greatly enhanced quantum repeater performance with atomic gases.
\end{abstract}

\maketitle

The distribution of entanglement over long distances is an
interesting challenge both for fundamental reasons and for
applications such as quantum key distribution and future
quantum networks. It is a difficult task because of
transmission losses, for example 1000 km of optical fiber
have a transmission of $10^{-20}$. Conventional
amplification as in classical telecommunications is ruled
out by the no-cloning theorem \cite{nocloning}. A possible
solution is the use of quantum repeaters \cite{briegel},
which are based on creating and storing entanglement in
moderate-distance elementary links and then extending it by
entanglement swapping. The DLCZ proposal \cite{DLCZ} for
realizing quantum repeaters has inspired many experiments
\cite{kuzmichexp,vanderwal,DLCZ-exp,yuan,SangouardRMP}. It
is based on ensembles of three-level systems, typically
atomic gases. The spontaneous Raman emission of a photon,
which we will call the {\it Stokes} photon, creates a heralded single atomic excitation in the ensemble. The detection of a
photon that could have come from either of the two
ensembles, in a way that erases all which-way information,
leads to an atomic excitation that is in a coherent
superposition state of being in either of the two
ensembles, cf. Fig. 1 of Ref. \cite{DLCZ}. This is an
entangled state, which forms the elementary link in the
DLCZ protocol. The entanglement can be extended by
re-converting the atomic excitations into {\it anti-Stokes}
photons and detecting them in the same way as the Stokes
photons.

The DLCZ protocol is attractive because it uses quite
simple ingredients. Unfortunately it is too slow to be
practical, even under optimistic assumptions for
re-conversion and detection efficiencies, storage times
etc. This recognition has recently led to proposals for
improvements \cite{improve,collins,P2M3,SangouardRMP}. The
most significant improvements can be achieved through the
use of multiplexing
\cite{collins,P2M3,SangouardRMP,kuzmichmulti}. We recently
proposed an attractive form of {\it temporal} multiplexing
that combines photon pair sources and quantum memories that
can store many temporal modes \cite{P2M3}. Solid-state
atomic ensembles are well suited for realizing such
temporal multi-mode memories
\cite{P2M3,CRIB,AFC,GEM,MMM-exp}. Nevertheless, given that
the most advanced experiments on quantum repeaters so far
have been performed with atomic gases \cite{DLCZ-exp,yuan},
it is also of great interest to search for ways of
achieving temporal multiplexing in such systems. In the
temporal multi-mode memories of Refs.
\cite{P2M3,CRIB,AFC,GEM,MMM-exp}, photons are stored on the
optical transition. The multi-mode character is achieved
thanks to the static inhomogeneous broadening of that
transition, in combination with photon echo techniques. In
the DLCZ protocol, which is based on Raman emission,
information is stored as a spin excitation. It is then
natural to look for a multi-mode protocol that involves
inhomogeneous broadening of the spin transition in
combination with spin echos. For other memory protocols
relying on spin echos see Refs. \cite{hetetOL,hosseini}.

In the DLCZ protocol, we are dealing with a large number
$N_A$ of Lambda atoms with two ground state levels $g$ and
$s$ and an excited state level $e$, cf. Fig. 1. Initially
all atoms are in $g$. The Raman emission of the Stokes
photon creates a state $|\psi\rangle=\frac{1}{\sqrt{N_A}}(
e^{i({\bf k}_w-{\bf k}_S)\cdot {\bf
x}_1}|s\rangle_1|g\rangle_2...|g\rangle_{N_A}+...+
e^{i({\bf k}_w-{\bf k}_S)\cdot {\bf
x}_{N_A}}|g\rangle_1...|g\rangle_{N_A-1}|s\rangle_{N_A})$,
where ${\bf x}_n$ is the position of the $n$-th atom, ${\bf
k}_w$ is the ${\bf k}$ vector of the write laser beam,
which is slightly detuned from the $g-e$ transition, and
${\bf k}_S$ that of the Stokes photon, which is emitted on
the $e-s$ transition. The state $|\psi\rangle$ describes a
single {\it spin wave} excitation with ${\bf k}$ vector
${\bf k}_w-{\bf k}_S$. Applying a read laser with ${\bf
k}_r=-{\bf k}_w$ on the $s-e$ transition will lead to the
emission of an anti-Stokes photon, associated with the
return to the atomic initial state
$|g\rangle_1...|g\rangle_{N_A}$. For the state
$|\psi\rangle$ the amplitude for the emission of the
anti-Stokes photon in direction ${\bf k}_{AS}$ is
proportional to $\sum \limits_{n=1}^{N_A} e^{-i({\bf
k}_{AS}+{\bf k}_S)\cdot {\bf x}_n}$, which for a large
ensemble consisting of many atoms is strongly peaked around
${\bf k}_{AS}=-{\bf k}_S$.

Now consider the case where there is inhomogeneous spin
broadening. This means that different atoms have slightly
different energy separations between $g$ and $s$, where for
each atom we will denote the detuning from the center
frequency of the $g-s$ transition by $\omega_n$. In this
case, if a time $t$ has elapsed since the Stokes emission,
the above state $|\psi\rangle$ has to be replaced by
$|\tilde{\psi}\rangle=\frac{1}{\sqrt{N_A}}( e^{-i \omega_1
t}e^{i({\bf k}_w-{\bf k}_S)\cdot {\bf
x}_1}|s\rangle_1|g\rangle_2...|g\rangle_{N_A}+...+ e^{-i
\omega_{N_A} t} e^{i({\bf k}_w-{\bf k}_S)\cdot {\bf
x}_{N_A}}|g\rangle_1...|g\rangle_{N_A-1}|s\rangle_{N_A})$,
and the anti-Stokes photon emission amplitude is now
proportional to $\sum \limits_{n=1}^{N_A} e^{-i \omega_n t}
e^{-i({\bf k}_{AS}+{\bf k}_S)\cdot {\bf x}_n}$, which in
general no longer has a peak for any direction of ${\bf
k}_{AS}$, due to the temporal phase factors $e^{-i\omega_n
t}$ that vary from atom to atom. However, suppose that we
are capable of flipping the sign of the atomic detunings,
$\omega_n \rightarrow -\omega_n$, for all $n$
\cite{CRIB,GEM,hetetOL,hosseini}. This requires the
inhomogeneous broadening to be controllable, for example it
could be due to a magnetic field gradient, cf. below. The
sign can then be flipped by flipping the sign of the
magnetic field \cite{hetetOL,hosseini}. In this case, after
another time $t$ all the temporal phases have canceled,
i.e. the state $|\tilde{\psi}\rangle$ has evolved back into
$|\psi\rangle$. If the read pulse is applied now, the
anti-Stokes emission will again be highly directional. If
no read pulse is applied, the spin wave will simply dephase
again.

We are now ready to describe the basic idea of our proposed
multi-mode protocol. For large $N_A$ and in the absence of
atom-atom interactions, spin waves corresponding to
different emission times and directions are completely
independent. Consider spin waves created at times $t_1,
t_2, t_3$ etc. If the atomic detunings are switched at a
time $T$, then these spin waves rephase at times $2T-t_1,
2T-t_2, 2T-t_3$ etc. By applying the read laser at a
specific time, one can ensure that only one specific spin
wave is in phase. Entanglement creation in the original
DLCZ protocol involves many unsuccessful attempts. Each
application of the write laser triggers the emission of a
Stokes photon in a given direction only with a small
probability (it has to be kept small because of
multi-excitation errors, see below). Most of the Stokes
photons will moreover be lost in long-distance
transmission. After every write pulse one has to wait for
information whether the photon was detected at the far-away
central station between the two repeater nodes under
consideration. For a typical distance between nodes of
$L_0=100$ km, the waiting time is $L_0/c=500 \mu$s, taking
into account the reduced speed of light in optical fibers.
This leads to low repetition rates, and thus to very low
entanglement creation rates. In contrast, Ref. \cite{P2M3}
showed that the capability to perform multi-mode storage
and selective recall allows one to apply the write laser
many times in quick succession. One can subsequently read
out exactly that spin wave for which the entanglement
generation was in fact successful and use it for
entanglement swapping etc. The present approach seems to
promise a greatly improved quantum repeater rate based on
this principle, see also sec. VI.A of Ref. \cite{ledingham}.

Unfortunately the described protocol has a serious problem,
which is absent for the scheme of Ref. \cite{P2M3}. In
typical experiments one detects only those Stokes photons
that are emitted in one specific direction, which is
defined by the geometry of the experiment. However, the
Stokes emission process itself is completely
non-directional (following a dipole emission pattern). This
means that most Stokes photons that are emitted go
undetected. But their emission is nevertheless associated
with the creation of unwanted spin wave excitations in $s$.
Suppose that the write laser was applied $N$ times,
defining $N$ separate time bins, and that a Stokes photon
was detected in the $k$-th time bin, leading to the
creation of a spin wave excitation. When reading out this
spin wave at some later time, there are several types of
contributions from the above-mentioned unwanted spin waves.

(1) There are spin waves associated with Stokes photons
that were emitted in the same ($k$-th) time bin. (a) Most
of these other Stokes photons will have been emitted in
other directions than the detected Stokes photon. When
reading out, the corresponding anti-Stokes photons will
therefore also be emitted in directions other than the
anti-Stokes photon that we are interested in. They thus
pose no problem for the protocol. (b) There is also the
possibility of emitting more than one Stokes photon during
the same time-bin in the same direction. Suppose that the
solid angle that is actually detected corresponds to a
fraction $\beta$ of all emitted Stokes photons, and that
the probability to emit a photon into this solid angle is
$p$. In typical experiments $\beta$ is in the range
$10^{-4}$ to $10^{-5}$ and $p$ in the range $10^{-2}$ to
$10^{-3}$. There is a probability $p^2$ to emit two photons
into the same solid angle, which implies that, given the
detection of a first photon, the conditional probability to
have a second, undetected one (which will lead to errors in
the protocol), is $2p$. The combinatorial factor of 2
arises because either of the photons in the two-photon
component of the state can lead to a detection. The
corresponding errors are well-known, they limit the value
of $p$ for the usual (single-mode) DLCZ protocol
\cite{DLCZ,SangouardRMP}.

(2) In our scenario with $N$ time bins there are additional errors due to the undetected Stokes photon emissions in all the other time bins. When reading out the spin wave associated with the
$k$-th time bin, the spin waves associated with all the
other time bins will not be in phase, as explained above.
They will thus not give rise to directional anti-Stokes
emission. However, the corresponding atoms in $s$ will
nevertheless be excited to $e$ by the read laser, the
resulting anti-Stokes emission will simply be
non-directional. With $p$ and $\beta$ as defined above, the
mean number of atoms transferred to the state $s$ during
the write process in each time bin is $\frac{p}{\beta}$, to first order in $p$ and summing over all directions of Stokes photon emission. (Note that typically $\frac{p}{\beta} \gg 1$ even though $p \ll 1$, cf. above). There are thus $\frac{(N-1)p}{\beta}$ atoms in $s$ which
correspond to out-of-phase spin waves. Only a fraction
$\beta$ of them will give rise to anti-Stokes photons that
are emitted exactly into the same solid angle as the
anti-Stokes photon that we are interested in. The total
error probability due to these out-of-phase spin waves is
thus $(N-1)p$.

Adding up the contributions from (1) and (2) gives a total
error probability of $(N+1)p$. There is then very little
advantage from using a multi-mode protocol. This is because
the choice of $p$ in a given repeater protocol is typically
determined by the size of the two-photon error. Suppose
that the acceptable error is $\epsilon$. (Its value depends
on the desired final fidelity of the repeater protocol and
on the number of repeater links, which depends on the
distance \cite{SangouardRMP}.) For an $N$-mode protocol one
then has to choose $p$ such that $(N+1)p=\epsilon$, or
$p=\frac{\epsilon}{N+1}$. On the other hand, the repeater
rate is proportional to $Np$ for a multi-mode protocol
\cite{P2M3}, i.e. it scales like $\frac{N}{N+1} \epsilon$,
which gives a modest improvement by a factor of two for
large $N$ compared to the single-mode case ($N=1$).

\begin{figure}[hr!]
{\includegraphics[width=1\columnwidth]{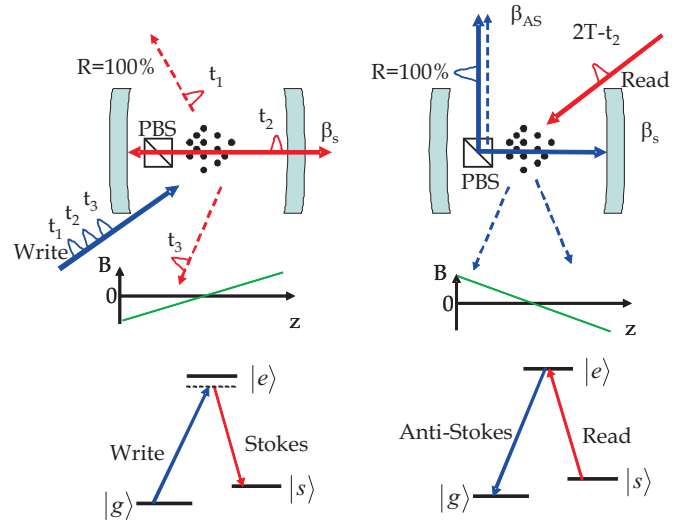}}
\caption{Temporally multiplexed creation and recall of spin
waves. Left: The atomic ensemble is excited by a sequence
of write pulses. Each write pulse can lead to the emission
of a Stokes photon. The cavity is in resonance with the
Stokes photons, such that emission into the cavity mode is
enhanced by a factor $F$, the cavity finesse. The cavity is
asymmetric, such that the Stokes photons leave the cavity
in one direction. Emission into the cavity occurs for a
fraction $\beta_S$ of all Stokes emissions, e.g. here at
time $t_2$. Every Stokes emission is associated with the
creation of a spin wave, which dephases due to the applied
magnetic field gradient. This dephasing can be reversed by
flipping the sign of the field, e.g. at time $T$. The spin
wave created by the Stokes emission at $t_2$ will be in
phase at $2T-t_2$. Applying a read pulse at this time
(right) creates an anti-Stokes photon, whose emission
direction is correlated with that of the Stokes photon due
to collective interference. A Stokes photon emitted into
the cavity creates a standing spin wave, the associated
anti-Stokes photon is therefore emitted into a
superposition of two counter-propagating modes. The
anti-Stokes photons have a polarization orthogonal to that
of the Stokes photons and are ejected from the cavity on
their first pass through the polarizing beam splitter
(PBS). (The mirrors are highly reflective for the
anti-Stokes photons.) Out-of-phase spin waves, e.g. those
created at $t_1$ and $t_3$, lead to the emission of
anti-Stokes photons without any preferred direction at
$t_2$. Since there is no cavity-induced enhancement for
them, only a small fraction $\beta_{AS} \ll \beta_S$ go in
the same direction as the ``good'' anti-Stokes photon which
is correlated to the Stokes photon from $t_2$.}
\label{writeread}
\end{figure}

We will now show that there is a way around this
disappointing conclusion. We focus on (2), i.e. the
anti-Stokes photons due to Stokes emissions in other time
bins, which is the main error mechanism in a multi-mode
protocol. The solution is to decrease the number of
unwanted spin waves that are created for every detected
Stokes photon, i.e. to increase the relative weight of the
detected Stokes photons compared to the non-detected ones.
However, this has to be done without a corresponding increase in the fraction
of non-directional anti-Stokes photons that are detected,
i.e. the detected fraction $\beta$ has to be much greater
for the Stokes than for the anti-Stokes photons, $\beta_S
\gg \beta_{AS}$. Under this condition, supposing that
$\beta_S$ is still much smaller than one, there are now
$(N-1)\frac{p}{\beta_S}$ atoms in $s$, where $p$ is still
the probability to emit a Stokes photon into the detected
spatial mode. Each atom in $s$ will lead to an anti-Stokes
photon emission into the detected mode with a probability
$\beta_{AS}$, giving a total error $(N-1)p
\frac{\beta_{AS}}{\beta_S}$, which is much smaller than
before if $\frac{\beta_{AS}}{\beta_S}\ll 1$. Adding the
error $2p$ from case (1) (i.e. from the same time bin), the
total error is now $p(2+(N-1)\frac{\beta_{AS}}{\beta_S})$.
As a consequence, if the acceptable error level is again
$\epsilon$, then the multi-mode repeater rate now scales
like
\begin{equation}
\frac{N \epsilon}{2+(N-1)\frac{\beta_{AS}}{\beta_S}},
\label{multimode}
\end{equation}
which tends towards $\epsilon \frac{\beta_S}{\beta_{AS}}$
for large $N$, compared to $\frac{\epsilon}{2}$ for $N=1$.
The multi-mode rate can thus be much greater than the
single-mode rate in this case.

We propose to increase $\beta_S$ without increasing
$\beta_{AS}$ by placing the atomic ensemble inside a cavity
that is in resonance with the Stokes transition, but that
is invisible for the anti-Stokes photons, cf. Fig. 1. Note
that it is not enough for the anti-Stokes photons to be
simply off-resonance with respect to the cavity, because in
that case they could not leave the cavity with high
probability, whereas a high collection efficiency for
``good'' anti-Stokes photons (those that are correlated
with the detected Stokes photons) is essential for a
successful repeater protocol. We propose to achieve
``invisibility'' of the cavity for the anti-Stokes photons
by having Stokes and anti-Stokes photons be at orthogonal
linear polarizations \cite{kuzmichexp}, see Fig. 1. A
cavity with finesse $F$ enhances the emission into one of
its modes by a factor of order $F$ compared to the
free-space situation, because the spectral density on
resonance is increased by $F$ \cite{purcell}. The cavity
concentrates the spectral density into a series of peaks of
width $\frac{c}{LF}$ separated by the free spectral range
$\frac{c}{L}$, where $L$ is the length of the cavity. In
the described situation we therefore have
$\frac{\beta_S}{\beta_{AS}}=F$ \cite{purcellfactor}. For
example, Refs. \cite{vuletic} and \cite{vuleticnph} had
$F=93$ and $F=240$ respectively for cavities containing
DLCZ-type atomic ensembles.

Such moderate-finesse cavities would already allow a great
enhancement in the quantum repeater rate, provided that the
number of time bins $N$ can be made sufficiently large; $N$
is directly determined by the size of the broadening. Since
the Stokes emission is an off-resonant Raman process, it
can in principle be made arbitrarily fast by choosing the
duration of the write pulse. However, the spin waves
corresponding to different write pulses will only be fully
distinguishable if there is complete dephasing of each spin
wave before the next write pulse. The duration of each time
bin thus has to be of order $\frac{1}{\gamma_{inh}}$, where
$\gamma_{inh}$ is the inhomogeneous width of the spin
transition. The other factor determining $N$ is the total
time available for emission, which in the context of
repeater protocols is given by the communication time
$L_0/c$, a typical value for which is 500 $\mu$s, cf.
above. The total number of modes would then be of order
$N\sim\frac{L_0}{c}\gamma_{inh}$=500 per MHz bandwidth.
There is no strong incentive to make $N$ much larger than
$F$, because the speedup thanks to multimode operation
begins to saturate at that level, cf. Eq.
(\ref{multimode}). In order to fully profit from a cavity
with $F$ of order 100 a broadening of order 1 MHz is thus
sufficient, with greater broadenings becoming relevant if
higher-finesse cavities are used. Typical gradients used in
magneto-optical traps ($\sim$ 10 G/cm) lead to Zeeman
broadenings for alkali atoms of order of a few MHz for mm
sized traps \cite{Felinto2005}. Spin rephasing can be
induced by reversing the current direction in the coils.
Storage and retrieval of light has been recently
demonstrated using the reversal of a magnetic field
gradient in a Rb vapor \cite{hetetOL,hosseini}. In Ref.
\cite{hosseini}, a controlled broadening of 1 MHz was
reversed in a few $\mu$s, which is fast enough for our
purposes (the switching time should be much shorter than
$L_0/c$).

The general requirements for implementing the DLCZ protocol
efficiently apply to the present proposal as well, in
particular one needs long storage times for the spin waves
and a high reconversion efficiency of spin waves into
anti-Stokes photons \cite{SangouardRMP}. Concerning the
latter, the cavity is of no assistance in the present case
since it is invisible to the anti-Stokes photons. However,
reconversion efficiencies as high as 50\% have already been
achieved for atomic ensembles in free space \cite{laurat}.
Long storage times can be achieved by placing the atomic
ensemble into an optical lattice. Light storage for 240 ms
has recently been demonstrated in a 3D lattice
\cite{Schnorrberger}, and single spin waves have been
stored for up to 8 ms in a 1D lattice
\cite{kuzmichlattice}. In the present context, the Stokes
photons couple to the cavity mode, leading to the creation
of a standing spin wave which is a superposition of two
plane waves with different values for $\Delta k=|{\bf
k}_w-{\bf k}_S|$. The spin wave with large $\Delta k$ will
decay faster because it is more sensitive to atomic motion
\cite{vuletic,bozhao}. In Ref. \cite{vuletic} the two decay
times differed by two orders of magnitude. Based on Ref.
\cite{Schnorrberger} this would still be compatible with ms
storage times for the fast-decaying component. Ref.
\cite{hau} recently stored light for over a second in a
Bose-Einstein condensate.

We have proposed a way of multiplexing the DLCZ quantum
repeater protocol using controlled reversible inhomogeneous
spin broadening in combination with moderate-finesse
cavities. Our approach, which can also be applied to
improved versions of the DLCZ protocol
\cite{SangouardRMP,improve}, opens a feasible avenue
towards greatly enhanced quantum repeater performance with
atomic gases.

We thank C. Ottaviani, N. Sangouard and N. Gisin for useful discussions, and we acknowledge support by the EU Integrated Project {\it Qubit Applications} and the Swiss NCCR {\it Quantum Photonics}.



\begin{thebibliography}{100}
\bibitem{purcellfactor} Note that it is not necessary for our purposes to have a
large Purcell factor, i.e. a substantial enhancement of the
overall Stokes emission rate, which would be achieved only
for $F \Omega > 1$, where $\Omega$ is the solid angle of
the cavity mode.
\end{thebibliography}
\end{document}